\documentclass[10pt,letterpaper]{article}
\usepackage[a4paper, margin=2.5cm]{geometry}

\usepackage{graphicx}
\usepackage{helvet}
\usepackage{authblk}
\usepackage[hidelinks, colorlinks=true, linkcolor=blue, urlcolor=blue, citecolor=blue]{hyperref}
\usepackage{amsmath} 
\usepackage{amssymb}
\usepackage{orcidlink} 
\usepackage[super,comma,sort&compress]  
   {natbib}
\usepackage{booktabs}
\usepackage{array}
\usepackage{amsmath}
\usepackage{nameref}
\usepackage{float}
\usepackage{orcidlink}
\makeatletter
\renewcommand{\maketitle}{\bgroup\setlength{\parindent}{0pt}
\begin{flushleft}
  \textbf{\@title}
  
  \@author
\end{flushleft}\egroup}
\makeatother

\title{{\fontsize{20pt}{20pt}\selectfont Passive aerodynamic robustness reduces disturbance amplification in flight}}

\author[1,\orcidlink{0000-0002-0509-2771}]{Lunbing Chen}
\author[1,\orcidlink{0000-0003-4982-5831}]{Suyang Qin} 
\author[1,\orcidlink{0009-0000-0931-3837}]{Qilin Wu}
\author[1,\orcidlink{0009-0009-1099-7426}]{Jinpeng Huang}
\author[1,\orcidlink{0009-0004-2932-4970}]{Yufei Yin}
\author[2, ]{Yong Chen}
\author[1,*,\orcidlink{0000-0002-1820-8622}]{Yang Xiang}
\author[1,,\orcidlink{0000-0001-9011-8309}]{Hong Liu}

\affil[1]{J. C. Wu Center for Aerodynamics, School of Aeronautics and Astronautics, Shanghai Jiao Tong University, Minhang, 200240, Shanghai, China}
\affil[2]{Commercial Aircraft Corporation of China, Ltd., Shanghai 200126, PR China}

\affil[*]{Lead Contact, Correspondence: \href{mailto:xiangyang@sjtu.edu.cn}{xiangyang@sjtu.edu.cn}}
\affil[**]{Correspondence: \href{mailto:hongliu@sjtu.edu.cn}{hongliu@sjtu.edu.cn}}

\begin{document}

\maketitle

\section*{SUMMARY}

Flight in turbulence is constrained not only by aerodynamic efficiency, but also by how strongly flow disturbances are transmitted into unsteady loads and dynamic responses. Although disturbance rejection is typically attributed to active control, birds often sustain fixed-wing gliding in disturbed air, suggesting that the wing itself may passively attenuate aerodynamic perturbations. 
Here, we show that avian wings reduce aerodynamic sensitivity to incoming disturbances. Compared with a geometrically matched airfoil wing, the avian wing exhibits lower lift-response gain, smoother stall transition, reduced force fluctuations, and a broader operative angle-of-attack range across turbulence intensities. These wing-level properties translate into an expanded passive stability envelope in rigid-flyer dynamics. 
Flow diagnostics indicate that this robustness is associated with delayed separation and redistribution of turbulent kinetic energy, which suppress large-scale flow instability and weaken disturbance transmission. This passive robustness comes at the cost of reduced aerodynamic efficiency, revealing an efficiency–robustness trade-off in disturbed flows. 
Our results identify aerodynamic sensitivity and control demand as essential metrics for flight performance in turbulence, and suggest passive aerodynamic robustness as a design principle for resilient flying systems.

\section*{KEYWORDS}

Passive aerodynamics, Disturbance sensitivity, Flight stability, Turbulent flow, Stability–efficiency trade-off, Avian flight

\section*{INTRODUCTION}
\label{sec:intro}

Flight in the atmosphere is shaped by a broad spectrum of flow structures, spanning relatively organized energy sources and unpredictable disturbances. Birds routinely encounter cyclones \cite{weimerskirch2019cyclone, lempidakis2022pelagic}, thermal updrafts \cite{weimerskirch2016frigate, kempton2022optimization, akos2008comparing}, orographic updrafts \cite{bishop2015roller, shamoun2016flap}, wave ridges \cite{richardson2011albatrosses, mohamed2022opportunistic}, shear flows \cite{sachs2005minimum, bousquet2017optimal, chen2026learning}, gusts \cite{fukami2023grasping, reynolds2014wing, cheney2020bird}, and turbulence \cite{tucker1972metabolism, ortega2014into, harvey2019wing, laurent2021turbulence}. Large-scale flow structures such as updrafts and shear flows are often sufficiently persistent to be exploited for energy harvesting during fixed-wing soaring and gliding \cite{weimerskirch2016frigate, bousquet2017optimal, mohamed2022opportunistic}. By contrast, small-scale gusts and turbulence are intermittent, broadband, and difficult to anticipate \cite{fukami2023grasping, tucker1972metabolism, ortega2014into, harvey2019wing}. These disturbances perturb the incoming velocity field and can be transmitted into unsteady aerodynamic forces, moments, and flight-dynamic responses, thereby challenging flight stability and, in extreme cases, survival \cite{reynolds2014wing, cheney2020bird}. Thus, flight in disturbed environments is governed not only by the ability to generate lift efficiently, but also by the sensitivity with which external perturbations are amplified into aerodynamic loads.

A well-documented strategy for coping with such disturbances is active wing control. Birds can rapidly adjust wing posture, camber, sweep, and flapping kinematics to regulate aerodynamic forces and moments. This disturbance-response control differs from slower morphing used to tune aerodynamic efficiency across flight speeds \cite{lentink2007swifts, cheney2021raptor}: it is transient, high-acceleration, and often deployed to reject perturbations. For example, steppe eagles perform rapid wing tucks that reduce load factor in turbulence \cite{reynolds2014wing}; bird wings can behave as suspension systems that buffer vertical gusts \cite{cheney2020bird}; and gulls modulate pitch stability through wing morphing in turbulent air \cite{harvey2019wing, harvey2022gull}. Flapping provides another active pathway for countering external perturbations \cite{ortega2014into}. These studies demonstrate that active wing actuation can effectively mitigate disturbance-induced loads and dynamic responses.

\begin{figure}[htbp]
    \centering
    \includegraphics[width=0.7\linewidth]{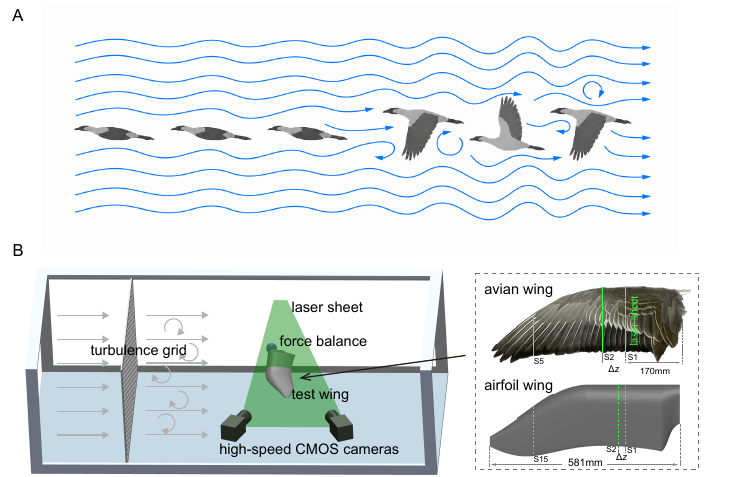}
    \caption{
    \textbf{Passive aerodynamic robustness as a mechanism for disturbance-resilient flight.}
    \textbf{(A)} Birds can actively respond to strong atmospheric disturbances through rapid wing or body adjustments, whereas under milder or persistent perturbations they may sustain fixed-wing gliding with limited actuation. This contrast raises the question of \textit{whether passive aerodynamic properties of the wing can attenuate disturbance transmission and reduce reliance on continuous active control.}
    \textbf{(B)} Experimental model problem used to isolate passive aerodynamic effects. Real avian wings and geometrically matched 3D-printed airfoil wings were exposed to controlled freestream turbulence generated by grids (more details in \autoref{subsec:method_aero} and \autoref{S_Fig:ExpSetup}). Laser-sheet planes mark the spanwise sections used for stereoscopic particle image velocimetry measurements of the suction-side flow field; adjacent sections were spaced by $\Delta z=70\,\mathrm{mm}$ and $\Delta z=20\,\mathrm{mm}$ for the five- and fifteen-section measurements, respectively.
    }
    \label{fig:modelProblem}
\end{figure}

However, active control may not be the only mechanism supporting flight stability in disturbed environments. Observations across species show that birds often sustain fixed-wing soaring, gliding, or landing under nonuniform and turbulent flow conditions with limited wing actuation. Albatrosses and falcons exploit shear flows and turbulent environments while maintaining largely gliding flight \cite{weimerskirch2016frigate, sachs2005minimum, mohamed2022opportunistic}, and near-ground turbulence does not preclude birds such as pigeons and swifts from gliding or landing with comparatively fixed wings \cite{lentink2007swifts, kleinheerenbrink2022optimization}. In many flight modes, active control occupies only a fraction of total flight time \cite{watkins2006atmosheric, wordley2008road, lempidakis2022estimating, weimerskirch2016frigate, reynolds2014wing, williams2020physical}. Migrating birds can even sustain flight during sleep-like states with reduced sensory engagement and intermittent control \cite{liechti2013first, rattenborg2016evidence}. These observations suggest that stable flight in disturbed air may also rely on passive physical mechanisms that attenuate disturbances before continuous high-frequency control is required (\autoref{fig:modelProblem}A).

This raises a central question: \textit{to what extent can passive aerodynamic properties reduce disturbance amplification and thereby contribute to flight stability in turbulent environments?}

Here, we address this question by isolating the role of passive aerodynamics under controlled turbulent inflow. We compare real avian wings with geometrically matched airfoil wings, allowing differences in disturbance response to be attributed primarily to aerodynamic and structural-surface properties rather than planform geometry. We show that avian wings exhibit reduced aerodynamic sensitivity to incoming disturbances, characterized by lower lift-response gain, smoother stall transition, reduced force fluctuations, and a broader operative angle-of-attack range across turbulence intensities. We then link these wing-level properties to system-level dynamics, showing that rigid flyers equipped with avian-wing aerodynamics maintain an expanded passive stability envelope. Flow diagnostics further indicate that this robustness is associated with delayed separation and redistribution of turbulent kinetic energy. Together, these results provide a quantitative framework for understanding how passive aerodynamic robustness can attenuate disturbance transmission, reduce reliance on active stabilization, and reshape the evaluation of flight performance in turbulent environments.

\section*{RESULTS}
\label{sec:results}

To isolate the role of passive aerodynamics in disturbance response, we compare real avian wings (swan goose, \textit{Anser cygnoides}) with geometrically matched rigid wings using an S1223 airfoil profile. Both configurations share the same planform geometry, ensuring that differences arise primarily from aerodynamic characteristics rather than shape (\autoref{fig:modelProblem}B).
Experiments were conducted in a wind tunnel at chord-based Reynolds numbers of $Re\sim O(10^5)$ and freestream turbulence intensities of $Tu=1.5\%$, $2.5\%$, and $7.4\%$. Time-resolved aerodynamic forces and boundary-layer flow fields were measured using a six-axis force balance and stereoscopic particle image velocimetry (SPIV) (\autoref{S_Fig:ExpSetup}). Detailed experimental procedures are provided in \autoref{sec:methods}.

\subsection*{Reduced aerodynamic sensitivity to turbulent disturbances}
\label{subsec:aeroChar}

\begin{figure}[htbp]
    \centering
    \includegraphics{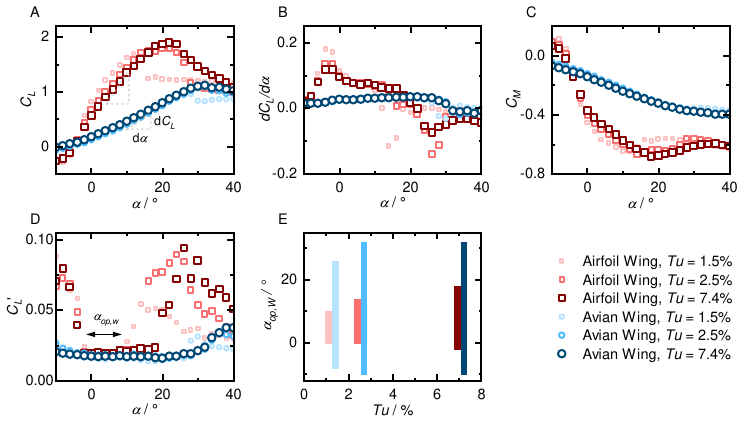}
    \caption{
    \textbf{Reduced aerodynamic sensitivity under turbulent disturbances.}
    Compared with the geometrically matched S1223 airfoil wing, the avian wing exhibits lower sensitivity to turbulent perturbations, characterized by a smaller lift-response gain, a more gradual stall transition, reduced pitching-moment sensitivity, lower lift fluctuations, and a broader operative angle-of-attack range across increasing turbulence intensities.
    (A) Lift coefficient $C_L$,
    (B) Lift-response gain quantified by $\mathrm{d}C_L/\mathrm{d}\alpha$,
    (C) Pitching-moment coefficient $C_M$, and 
    (D) Standard deviation of lift coefficient $C'_L$ as a function of $\alpha$.
    (E) Operative angle-of-attack range $\alpha_{\mathrm{op,w}}$, defined as the range of $\alpha$ over which lift fluctuations remain bounded (see also \autoref{S_Fig:liftForce_PSD}).
    Here, $Tu=1.5\%$ represents clean flow, whereas $Tu=2.5\%$ and $7.4\%$ denote moderate and high freestream turbulence generated by the grid.
    Additional aerodynamic data at a higher Reynolds number are provided in \autoref{S_Fig:force_swan_V15}.
    }
    \label{fig:forceCoe_V11}
\end{figure}

\noindent
\textit{Can passive aerodynamics reduce the sensitivity of a wing to turbulent disturbances, thereby contributing to stability without active morphing?} To examine this question, we first evaluate how the aerodynamic response varies with angle of attack ($\alpha$) and freestream turbulence intensity. The time-averaged and fluctuating aerodynamic characteristics of the avian wing and the matched S1223 airfoil wing are compared in \autoref{fig:forceCoe_V11}. Despite close geometric matching (\autoref{fig:modelProblem}B), the two wings exhibit systematically different responses to variations in $\alpha$ and $Tu$.

A first distinction lies in the lift response. The airfoil wing exhibits a larger lift slope ($\mathrm{d}C_L/\mathrm{d}\alpha$) and a pronounced peak in $C_L$ at a relatively low angle of attack, followed by a rapid post-stall decrease. In contrast, the avian wing shows a smaller lift-response gain and sustains lift over a broader range of $\alpha$, with delayed stall onset and a more gradual post-stall variation (\autoref{fig:forceCoe_V11}A,B).
Atmospheric disturbances perturb the incoming velocity field, inducing variations in the effective angle of attack and airspeed \cite{fukami2023grasping, reynolds2014wing, cheney2020bird, ortega2014into}. Under a quasi-steady approximation, the resulting lift variation can be expressed as $\Delta C_L \approx (\mathrm{d}C_L/\mathrm{d}\alpha)(w_\mathrm{gust}/u_\infty)$ \cite{nakata2015cfd, wang2016predictive}, where $w_\mathrm{gust}$ is the vertical gust velocity fluctuation and $u_\infty$ is the freestream velocity magnitude. The lift-response gain $\mathrm{d}C_L/\mathrm{d}\alpha$ therefore governs the amplification of external perturbations, such that a smaller slope leads to weaker transmission of disturbances into aerodynamic loads.

A similar difference is found in the pitching response. The pitching-moment coefficient ($C_M$) of the airfoil wing varies more rapidly with $\alpha$, whereas the avian wing exhibits a less negative and more gradual dependence on $\alpha$ (\autoref{fig:forceCoe_V11}C). This reduced moment sensitivity suggests a smaller aerodynamic tendency for disturbance-induced changes in pitch, which is examined further in the following \autoref{subsec:stability}.

The fluctuating loads further distinguish the two wings. The airfoil wing exhibits larger lift fluctuations ($C'_L$), particularly at elevated turbulence intensities, whereas the avian wing maintains comparatively low fluctuation levels over the measured range of $\alpha$ (\autoref{fig:forceCoe_V11}D). Consistent with this behavior, the avian wing retains a broader operative angle-of-attack range, $\alpha_{\mathrm{op,w}}$, under all tested turbulence conditions (\autoref{fig:forceCoe_V11}E, \autoref{S_Fig:liftForce_PSD}). For example, at $Tu=7.4\%$, the operative range of the avian wing extends from $-10^\circ$ to $34^\circ$, whereas that of the airfoil wing is limited to $-2^\circ$ to $18^\circ$.

Taken together, these results show that the avian wing exhibits lower aerodynamic sensitivity to turbulent perturbations, as reflected by smaller lift-response gain, weaker pitching-moment sensitivity, lower force fluctuations, and a broader usable range of $\alpha$. These wing-level characteristics indicate that passive aerodynamic properties can attenuate the transmission of external disturbances into aerodynamic load variations.

\subsection*{Translation of reduced aerodynamic sensitivity into system-level stability}
\label{subsec:stability}

\noindent
\textit{Does the reduced aerodynamic sensitivity observed at the wing level translate into improved stability at the system level?}
To establish this connection, we extend the analysis from wing-level aerodynamics to the longitudinal dynamics of the full flyer. 
We model the longitudinal dynamics of rigid flyers using small-perturbation theory \cite{bossert2003introduction}. Two configurations are considered: one equipped with the avian wing and the other with the S1223 airfoil. To isolate aerodynamic effects, both flyers share identical inertial properties derived from anatomical data \cite{harvey2022gull, harvey2022birds}. The aerodynamic stability derivatives are obtained directly from the wind tunnel measurements described above. The resulting linearized model is summarized in \autoref{subsec:method_stability}.

\begin{figure}[htbp]
    \centering
    \includegraphics[width=1.0\linewidth]{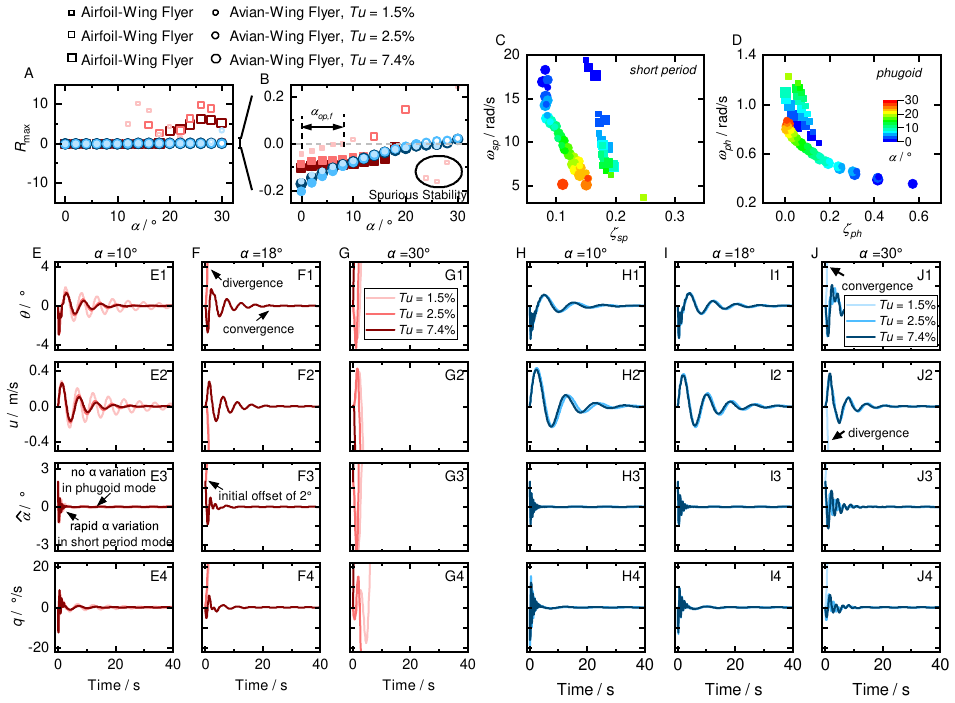}
    \caption{
    \textbf{Expanded passive stability envelope at the system level.}
    The avian-wing flyer maintains passive stability over a broader range of conditions than the airfoil-wing flyer across varying turbulence intensities.
    (A,B) Maximum real part of the eigenvalues, $\mathrm{Re}_{\max}$, as a function of angle of attack $\alpha$, where $\mathrm{Re}_{\max}<0$ denotes stable flight. The corresponding operative stability range is denoted $\alpha_{\mathrm{op,f}}$.
    (C,D) Natural frequencies ($\omega$) and damping ratios ($\zeta$) of the short-period and phugoid modes. Squares and circles denote the airfoil-wing and avian-wing configurations, respectively; marker size indicates turbulence intensity ($Tu=1.5\%, 2.5\%, 7.4\%$), and color indicates trim angle of attack.
    (E--J) Time-domain responses to an initial $2^\circ$ perturbation in angle of attack. Convergent responses indicate passive stability, whereas divergence indicates instability.
    At $\alpha \gtrsim 24^\circ$ for the airfoil-wing flyer at $Tu=1.5\%$, the apparent re-stabilization is a non-physical artifact of the linear small-perturbation model and is excluded from the analysis.
    }
    \label{fig:stability}
\end{figure}

The stability characteristics are evaluated by examining the maximum real part of the eigenvalues, denoted $\mathrm{Re}_\mathrm{max}$, over a range of trim angles of attack ($0^\circ$ to $30^\circ$). According to linear stability theory, the system is stable if and only if $\mathrm{Re}_\mathrm{max} < 0$ \cite{bossert2003introduction}. As shown in \autoref{fig:stability}A,B, the airfoil-wing flyer exhibits a limited range of passive stability, remaining stable only below $\alpha = 10^\circ$ at $Tu=1.5\%$, $\alpha = 14^\circ$ at $Tu=2.5\%$, and $\alpha = 18^\circ$ at $Tu=7.4\%$. In contrast, the avian-wing flyer maintains $\mathrm{Re}_\mathrm{max} < 0$ over a substantially broader range of conditions, with instability appearing only near the upper bound of $\alpha$ under low turbulence. The larger $\alpha_{\mathrm{op,f}}$ observed for the avian-wing configuration indicates that the reduced aerodynamic sensitivity identified at the wing level translates directly into an expanded passive stability envelope.

The modal properties of the system clarify the nature of this stability. The natural frequencies ($\omega$) and damping ratios ($\zeta$) (defined in \autoref{subsec:method_stability}) associated with the short-period and phugoid modes are shown in \autoref{fig:stability}C,D. The avian-wing flyer exhibits a lower damping ratio in the short-period mode, indicating relatively responsive attitude dynamics. In contrast, in the phugoid mode, it consistently exhibits a higher damping ratio and lower oscillation frequency compared to the airfoil-wing configuration, with weak dependence on turbulence intensity. These characteristics correspond to a slower but more strongly damped long-period response, which promotes passive recovery toward equilibrium following disturbances.

The time-domain responses to a representative perturbation illustrate these differences. The response to an initial $2^\circ$ offset in angle of attack is shown in \autoref{fig:stability}E–J. For conditions within the stable regime, the dynamics exhibit a short-period transient followed by a longer-period phugoid motion. The avian-wing flyer displays sustained bounded responses across a wider range of conditions, whereas the airfoil-wing flyer exhibits divergence at higher $\alpha$. For example, the airfoil-wing configuration diverges at $\alpha = 18^\circ$ under low and moderate turbulence and at $\alpha = 30^\circ$ for all turbulence levels, while the avian-wing configuration remains stable under most tested conditions, except at the extreme case of $\alpha = 30^\circ$ with minimal turbulence.

Overall, these results demonstrate that the reduced aerodynamic sensitivity identified at the wing level translates into a systematically expanded passive stability envelope at the system level. The avian-wing configuration attenuates the transmission of external disturbances into destabilizing dynamic responses, thereby reducing the reliance on active control for maintaining stable flight.

\subsection*{Flow--structure interactions underlying passive aerodynamic robustness}
\label{subsec:flowAnalysis}

\noindent
\textit{What physical mechanisms underlie the reduced aerodynamic sensitivity and the resulting stability observed above?}
To address this question, we examine the flow fields over the avian and airfoil wings using SPIV \cite{qiu2021characteristics, qin2025flow, wu2025response}. Measurements were acquired at multiple spanwise sections (\autoref{fig:modelProblem}B). Although the two configurations are closely matched in their macroscopic geometric parameters, the avian wing retains biological surface and structural features, including roughness, compliance, and permeability, that are absent in the engineered airfoil \cite{zhou2021effect, matloff2020how}. Throughout this analysis, the avian wing is treated as an integrated aerodynamic system. Definitions of the relevant flow quantities are provided in \autoref{sub:method_flow}.

\begin{figure}[htbp]
    \centering
    \includegraphics[width=1.0\linewidth]{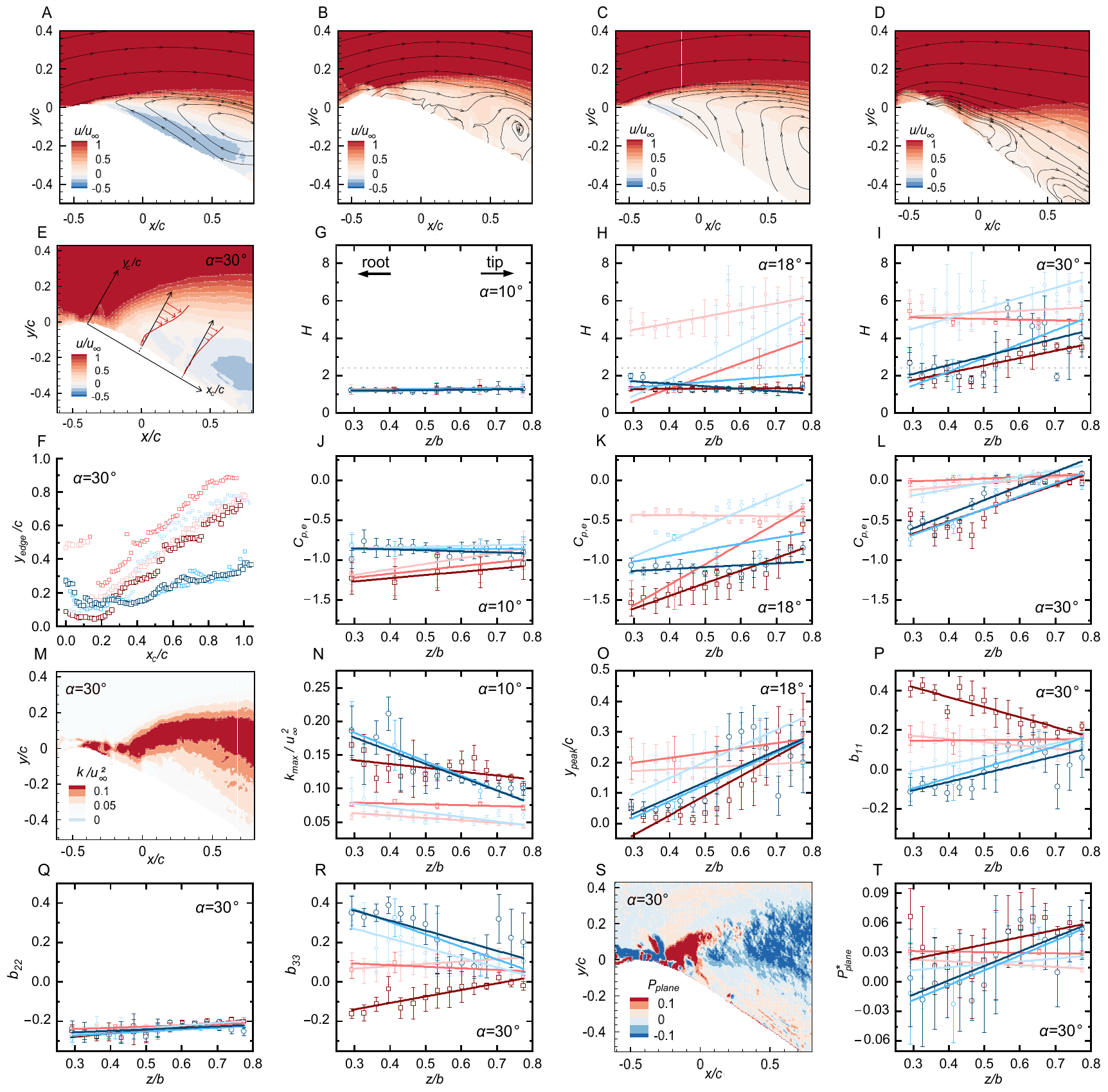}
    \caption{
    \textbf{Flow--structure interactions associated with delayed separation and redistributed turbulent activity over the avian wing.}
    Definitions of the reported flow quantities are provided in \autoref{sub:method_flow}.
    \textbf{(A--D)} Normalized time-averaged velocity fields, $\bar{u}/u_\infty$ at $z/b=0.3$ for $Tu=1.5\%$: \textbf{(A)} airfoil wing at $\alpha=18^\circ$, \textbf{(B)} avian wing at $\alpha=18^\circ$, \textbf{(C)} airfoil wing at $\alpha=30^\circ$, and \textbf{(D)} avian wing at $\alpha=30^\circ$.
    \textbf{(E)} Wall-normal velocity profiles used to identify the boundary-layer edge.
    \textbf{(F)} Chordwise distribution of normalized boundary-layer thickness, $y_{\mathrm{edge}}/c$, at mid-span.
    \textbf{(G--I)} Spanwise distributions of the chordwise-averaged shape factor, $H$, showing more limited separation over the avian wing (see also \autoref{S_Fig:flow_airfoil} and \autoref{S_Fig:flow_avian}).
    \textbf{(J--L)} Spanwise distributions of the chordwise-averaged edge-pressure coefficient, $C_{p,e}$, estimated from the boundary-layer edge velocity at $\alpha=10^\circ$, $18^\circ$, and $30^\circ$; more negative values indicate stronger suction over the upper surface.
    \textbf{(M)} Distribution of normalized turbulent kinetic energy, $k/u_\infty^2$, at $\alpha=30^\circ$.
    \textbf{(N)} Spanwise distribution of the chordwise-averaged TKE peak magnitude, $k_{\mathrm{max}}/u_\infty^2$.
    \textbf{(O)} Spanwise distribution of the chordwise-averaged wall-normal location of the TKE peak, $y_{\mathrm{peak}}/c$.
    \textbf{(P--R)} Spanwise distributions of the chordwise-averaged Reynolds-stress anisotropy components, including the streamwise component $b_{11}$, wall-normal component $b_{22}$, and spanwise component $b_{33}$ (\autoref{eq:bii}), indicating a stronger spanwise fluctuation component over the avian wing.
    \textbf{(S)} Contours of normalized in-plane TKE production, $P_{\mathrm{plane}}$.
    \textbf{(T)} Spanwise distribution of the chordwise-averaged wall-normal integral of normalized in-plane TKE production, $P_{\mathrm{plane}}^{*}$ (\autoref{eq:P_plane}).
    }
    \label{fig:flow_analysis}
\end{figure}

\autoref{fig:flow_analysis}A--D compares representative time-averaged velocity fields and streamline patterns at mid-span. 
At $\alpha=10^\circ$, both wings exhibit predominantly attached flow, consistent with the linear lift regime in \autoref{fig:forceCoe_V11}A. 
As $\alpha$ increases to $30^\circ$, both wings develop separated regions.
The spanwise-resolved velocity fields in \autoref{S_Fig:flow_airfoil} and \autoref{S_Fig:flow_avian} further show that, particularly under low-turbulence conditions, the avian wing delays the development of large-scale separation and confines low-speed or reversed-flow regions to a smaller portion of the suction-side flow field. 
At higher freestream turbulence intensities, enhanced incoming disturbances introduce additional mixing and unsteadiness, which partially reduce the contrast between the avian and airfoil wings \cite{schlichting2017boundary, thompson2023effects}. 
Therefore, the reduced separation observed over the avian wing under low-turbulence conditions more directly reflects the intrinsic passive aerodynamic characteristics of the wing.

This separation state can be quantified using the boundary-layer thickness and the shape factor $H$ \cite{schlichting2017boundary}. At $\alpha=30^\circ$, the avian wing exhibits a smaller boundary-layer thickness than the airfoil, with comparatively weak dependence on turbulence intensity (\autoref{fig:flow_analysis}G). The spanwise distributions of $H$ (\autoref{fig:flow_analysis}G--I) indicate widespread separation over the airfoil, whereas the avian wing preserves attached or only partially separated flow. These trends are consistent with the broader operative angle-of-attack range identified in \autoref{fig:forceCoe_V11}.

The implications of these differences are reflected in the surface pressure distribution. As shown in \autoref{fig:flow_analysis}(E,F), the velocity fields are analyzed in a chord-fixed coordinate system $(x_c/c,y_c/c)$, and the boundary-layer edge velocity $u_e$ is used to estimate the Bernoulli-estimated edge pressure coefficient via $C_{p,e} = 1 - (u_e/u_\infty)^2$ \cite{schlichting1961boundary}. This approximation should be interpreted as an edge-pressure estimate and is used here primarily for comparative analysis across the two wings.
The spanwise distributions of $C_{p,e}$ (\autoref{fig:flow_analysis}J--L) show that, at moderate to high $\alpha$, the airfoil undergoes a substantial loss of suction near the tip of the span, consistent with large-scale separation (\autoref{fig:flow_analysis}G--I), whereas the avian wing retains suction over a larger spanwise extent especially at $\alpha=30^\circ$.

To examine why the two wings differ in their separation behavior, we next consider the turbulent kinetic energy (TKE) field \cite{pope2000turbulent} (\autoref{fig:flow_analysis}M). At $\alpha=30^\circ$, the two configurations exhibit comparable values of the normalized TKE peak magnitude, $k_{\mathrm{max}}/u_\infty^2$ (\autoref{fig:flow_analysis}N), indicating that the principal difference does not lie in the overall level of turbulent activity alone. Instead, a key distinction is found in the wall-normal position of the TKE peak, $y_{\mathrm{peak}}/c$ (\autoref{fig:flow_analysis}O). Over the avian wing, the TKE peak remains closer to the surface across the measured span and across turbulence conditions, indicating that turbulent energy penetrates more effectively into the near-wall region. This spatial distribution is consistent with enhanced momentum exchange within the boundary layer, which helps replenish near-wall low-momentum fluid under an adverse pressure gradient and thereby delays separation \cite{schlichting2017boundary, pope2000turbulent}.

The Reynolds-stress anisotropy provides additional information on the structure of the separated flow \cite{pope2000turbulent, lee2015direct}. As shown in \autoref{fig:flow_analysis}P--R, the airfoil flow remains more strongly weighted toward streamwise fluctuations ($b_{11}$), whereas the avian wing exhibits a larger spanwise fluctuation component ($b_{33}$). This shift does not, by itself, identify a specific coherent structure, but it does indicate a more strongly three-dimensional separated flow. Such enhanced three-dimensionality is consistent with greater spanwise momentum redistribution and reduced spanwise coherence of the separated shear layer.

We further assess mean-to-turbulence energy transfer using the in-plane resolved TKE production field \cite{pope2000turbulent}. As shown in \autoref{fig:flow_analysis}S,T, the airfoil exhibits predominantly positive production, consistent with sustained transfer of energy from the mean flow into the separated shear layer \cite{orlandi2019turbulent}. By contrast, the avian wing exhibits a more spatially heterogeneous production field, including locally negative regions. Although these negative values do not by themselves establish a complete energy budget, they are consistent with more complex local energy redistribution, including possible backscatter. This pattern suggests that turbulent activity over the avian wing is less spatially concentrated and less persistently amplified within the separated region, which is compatible with the weaker large-scale separation observed above \cite{balin2021direct}.

Taken together, these results indicate that the passive robustness of the avian wing arises from coupled flow--structure interactions that alter both the location and the directional distribution of velocity fluctuations. Relative to the airfoil, the avian wing promotes stronger near-wall momentum exchange, delays the development of large-scale separation, and exhibits a more three-dimensional separated flow.

\subsection*{A trade-off between aerodynamic efficiency and passive robustness}
\label{subsec:tradeoff}

\noindent
\textit{What is the cost of passive aerodynamic robustness, and what does this imply for aerodynamic design in turbulent environments?}
We next examine the energetic implications of passive aerodynamic robustness by considering two components: the base aerodynamic cost and the dynamic control expenditure.

\begin{figure}[htbp]
    \centering
    \includegraphics{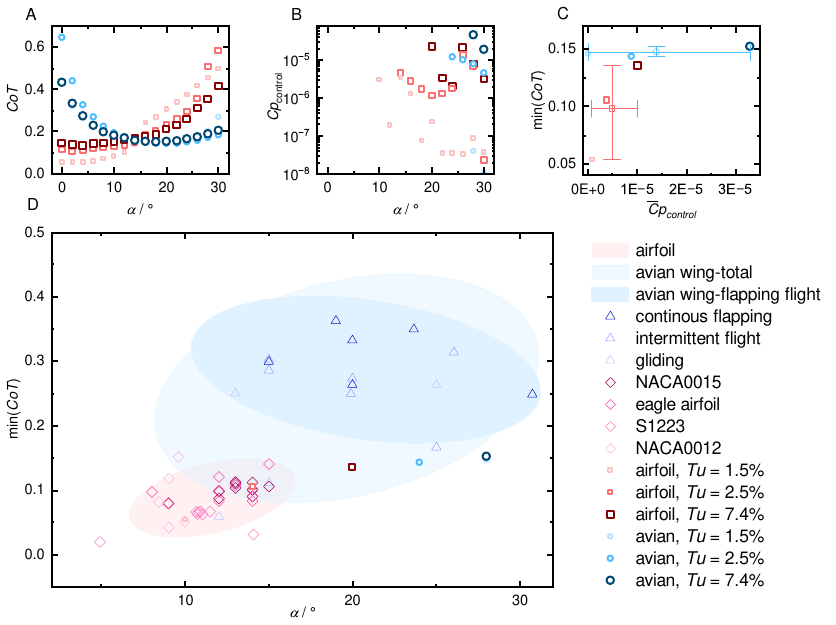}
    \caption{
    \textbf{Energy--stability trade-off across aerodynamic and control metrics.}
    (A) Base aerodynamic cost, quantified by the Cost of Transport ($CoT$), as a function of angle of attack $\alpha$ under varying turbulence intensities. 
    (B) Dynamic control expenditure, represented by the average control-power coefficient $\overline{C}_{P,\mathrm{control}}$. 
    (C) Combined mapping of minimum aerodynamic cost $\min(CoT)$ versus $\overline{C}_{P,\mathrm{control}}$. 
    (D) Global performance envelope showing $\min(CoT)$ versus the maximum operative angle-of-attack range of the wing, $\max(\alpha_{\mathrm{op,w}})$. Engineered airfoils (pink) and avian wings (blue) are shown with shaded 90\% confidence ellipses. Literature data are compiled from Ref. \cite{withers1981aerodynamic, lees2016influence, yap2001effect, cao2011performance, tsuchiya2013influence} for $Tu \in [0.5\%, 9.5\%]$.
    }
    \label{fig:costAnalysis}
\end{figure}

The base cost is characterized by the Cost of Transport ($CoT = C_D/C_L$), which reflects steady-state aerodynamic efficiency \cite{pennycuick2008modelling}. As shown in \autoref{fig:costAnalysis}A, the airfoil wing exhibits lower $CoT$ over the low-angle-of-attack regime, indicating higher efficiency under nominal conditions. 
The dynamic control expenditure required to stabilize the system under disturbances is quantified by the average control-power coefficient $\overline{C}_{P,\mathrm{control}}$ (derived in \autoref{subsec:method_control}). As shown in \autoref{fig:costAnalysis}B and \autoref{S_Fig:cost}, for comparable turbulence intensities, the avian-wing configuration requires higher control effort when active stabilization is engaged.
A combined view of these metrics is presented in \autoref{fig:costAnalysis}C, where $\min(CoT)$ is plotted against $\overline{C}_{P,\mathrm{control}}$. The airfoil occupies a region of lower aerodynamic and control cost, whereas the avian wing is shifted toward higher values in both dimensions. Within this framework, the avian wing does not minimize energetic expenditure.

However, this energetic comparison alone does not capture the functional advantage of the avian configuration. 
Owing to its extended operative and passive stability envelopes (\autoref{fig:forceCoe_V11}E, \autoref{fig:stability}A,B), the avian-wing flyer can maintain stable flight over a substantially wider range of angles of attack without active control. 
\autoref{fig:costAnalysis}D maps the minimum aerodynamic cost of transport, $\min(CoT)$, against the maximum operative angle-of-attack range of the wing, $\max(\alpha_{\mathrm{op,w}})$, combining the present measurements with literature data \cite{withers1981aerodynamic, lees2016influence, yap2001effect, cao2011performance, tsuchiya2013influence}. 
Engineered airfoils cluster in a regime characterized by low $\min(CoT)$ and limited $\max(\alpha_{\mathrm{op,w}})$, consistent with designs optimized primarily for steady-state aerodynamic efficiency. 
In contrast, avian wings occupy a distinct regime with higher $\min(CoT)$ but substantially larger $\max(\alpha_{\mathrm{op,w}})$, indicating a broader range of conditions over which aerodynamic loads remain bounded and passive flight can be maintained without continuous active stabilization.

Taken together, these results reveal a trade-off between aerodynamic efficiency and passive robustness. 
Although the avian wing incurs a higher steady aerodynamic cost, it enables stable flight over a broader disturbance envelope with reduced reliance on continuous control input. 
In turbulent environments, where disturbances are persistent and can occur on time scales comparable to vehicle dynamics, the reduction in active stabilization demand may partly offset the penalty in steady aerodynamic efficiency. 
These findings suggest that aerodynamic design for disturbed flows should be evaluated beyond steady-state efficiency alone, incorporating disturbance sensitivity, control demand, and passive robustness as key metrics for expanding the operational envelope.

\section*{DISCUSSION}

Atmospheric disturbances continuously perturb flight dynamics, yet birds are able to maintain stable fixed-wing gliding without continuous high-frequency active control. In this study, we combined aerodynamic measurements, flow diagnostics, stability analysis, and energetic evaluation to investigate this capability. The results consistently show that avian wings exhibit reduced aerodynamic sensitivity to perturbations, which translates into an expanded passive stability envelope at the system level. The underlying flow mechanism involves delayed separation and redistribution of turbulent energy, which attenuate the transmission of external disturbances into aerodynamic loads and dynamic responses.

A central outcome of this work is the identification of a trade-off between aerodynamic efficiency and passive robustness. Conventional aerodynamic design prioritizes maximizing lift-to-drag ratio ($L/D$) to reduce steady-state energy expenditure. However, our results show that configurations achieving high peak efficiency are also more sensitive to disturbances, leading to a restricted stable operating range and increased reliance on active control. In contrast, the avian wing exhibits higher baseline aerodynamic cost but maintains stability over a substantially broader range of conditions. This indicates that the practical performance of a wing in turbulent environments may not be determined solely by its peak efficiency, but also by its sensitivity to perturbations and its ability to sustain stable flight without continuous control input.

This trade-off can also be interpreted in the context of morphology facilitating control \cite{blickhan2007intelligence, muller2017what}. The aerodynamic characteristics of the avian wing effectively reduce the need for active stabilization by attenuating disturbances at the physical level. From a system perspective, this suggests a layered control structure: passive aerodynamic properties handle a portion of disturbance rejection under typical conditions, while active control is engaged only when disturbances exceed this passive capability. Such a division reduces the frequency and magnitude of control intervention required during flight.

The implications for aerodynamic design can be significant. Current autonomous aerial systems operating in turbulent environments rely heavily on sensor–actuator feedback loops \cite{floreano2015science, kleinheerenbrink2017multi, thompson2023effects, mohamed2014fixed, harvey2023lessons}, which introduce additional complexity, weight, and energy consumption. The present results suggest an alternative design strategy: incorporating passive aerodynamic robustness into the wing itself to reduce disturbance sensitivity. By partially shifting the stabilization burden from control systems to the mechanical structure \cite{reynaga2026mechanical}, it may be possible to reduce overall system complexity while maintaining or improving operational reliability.

\subsection*{Limitations of the study}
\label{subsec:limitations}

Several limitations of the present study should be noted. First, the analysis is based on an isolated fixed-wing configuration and focuses on longitudinal perturbations. Real atmospheric disturbances are inherently three-dimensional and time-dependent, and their interaction with full-body dynamics may introduce additional effects not captured here. Second, the quasi-steady framework adopted in the stability analysis may not fully represent the response to large, unsteady gusts. Third, although the avian wing is treated here as an integrated system, the specific contributions of individual morphological features—such as surface roughness, structural flexibility, or permeability—remain to be isolated in future work.

Finally, the present results provide an indirect assessment of the energetic implications of passive stability. Direct validation using freely flying systems, either through biologging of natural flyers or controlled experiments with bio-inspired aerial vehicles, will be necessary to quantify the extent to which reduced control demand translates into improved overall efficiency in realistic environments.

\section*{METHODS}
\label{sec:methods}

\subsection*{Experimental setup}
\label{subsec:method_expSetup}

\noindent
Experiments were conducted in a closed-loop wind tunnel at Shanghai Jiao Tong University, with a test section of $1.2 \times 0.9 \times 3.0~\mathrm{m}^3$ \cite{qiu2021characteristics, qin2025flow, wu2025response}. Freestream velocities $u_\mathrm{\infty}$ of $11$ and $15~\mathrm{m/s}$ were used, corresponding to chord-based Reynolds numbers $Re=u_\infty\bar{c}/\nu$ of $1.2\times10^5$ and $1.6\times10^5$, where $\bar{c}$ is the mean chord length and $\nu$ is the kinematic viscosity of air.

Freestream turbulence was generated using grids (\autoref{S_Fig:ExpSetup}B) \cite{broatch2022automatized}, yielding three conditions: $Tu=1.5\%$ (no grid), $2.5\%$, and $7.4\%$. $Tu=\sigma_u/u_\infty$ is the turbulence intensity, measured using hot-wire anemometry \cite{wu2025response}, where $\sigma_u$ is the root-mean-square of the streamwise velocity fluctuation. These values are representative of atmospheric turbulence levels ($2\%$–$8\%$) reported in previous field measurements \cite{watkins2006atmosheric, wordley2008road}.
For each condition, turbulence intensity was measured upstream of the wing near the leading edge plane. Velocity time series were sampled at $1\ \mathrm{kHz}$, and the one-sided power spectral density was obtained via fast Fourier transform. The spectra were normalized by the freestream velocity $u_\infty$ and the characteristic grid length $M$ (\autoref{S_Fig:Tu}).

A swan goose (\textit{Anser cygnoides}) wing, obtained from natural casualties at a breeding facility (Suzhou, China) \cite{huang2025geese}, was fixed in a gliding posture following \cite{harvey2019wing, harvey2021gull}, characterized by the elbow angle of $\approx150^\circ$, the manus angle of $\approx180^\circ$, and the sweep angle of $ \approx34^\circ$, with mean chord length $\overline{c} = 0.163 \, \mathrm{m}$, semi-span $b=0.581~\mathrm{m}$ and semi-aspect ratio $AR = 3.57$. A geometrically matched rigid wing with an S1223 airfoil section was fabricated for comparison (\autoref{fig:modelProblem}B) \cite{shyy1999flapping, usherwood2003aerodynamics, maeng2013modeling, rader2023morphological}. Both models share identical planform geometry, isolating the effects of surface morphology and aeroelastic properties \cite{matloff2020how, favier2009passive}.

\subsection*{Aerodynamic and flow measurements}
\label{subsec:method_aero}

\noindent
Aerodynamic forces and moments were measured using a six-axis force balance (ATI Gamma), sampled at $1~\mathrm{kHz}$ for $30~\mathrm{s}$ per condition. Data were low-pass filtered at $40~\mathrm{Hz}$, and each case was repeated ten times. 
Force coefficients ($C_L, C_D, C_M$) and their standard deviations were computed.
The deviation between different runs was less than $0.7\%$ \cite{qin2025flow}. 

Flow fields were measured using stereo particle image velocimetry (SPIV), providing three-component velocity fields in streamwise planes \cite{qiu2021characteristics, qin2025flow}. For each condition, 300 image pairs were acquired at $200~\mathrm{Hz}$, with $50 \, \mathrm{\mu s}$ interval between two laser pulses. Measurements were performed at multiple spanwise locations, with higher spatial resolution in cases exhibiting flow separation (\autoref{fig:modelProblem}B). The resulting percentage of good vectors was over $96\%$ using the peak ratio method \cite{charonko2013estimation, chen2026load}.

More details about measurement can be found in Ref. \cite{qiu2021characteristics, qin2025flow, chen2026load}.

\subsection*{Dynamic stability analysis}
\label{subsec:method_stability}

\noindent
Longitudinal dynamics were modeled using a linearized state-space formulation \cite{bossert2003introduction, harvey2022gull}:
\begin{equation}
    \dot{\mathbf{x}} = A\mathbf{x}, \qquad \mathbf{x}=[\theta, u_\mathrm{f}, \alpha, q]^T,
\end{equation}
where the system matrix $A$ was constructed from aerodynamic derivatives obtained from measured force data, $\theta$ is the pitch angle, $u_\mathrm{f}$ is the forward speed, $\alpha$ is the angle of attack, and $q=\dot{\theta}$ is the pitch rate.

Stability was assessed from the eigenvalues $\lambda$ of $A$, with passive stability defined by $\mathrm{Re}(\lambda)<0$. The passive stability envelope was evaluated over $\alpha \in [0^\circ,30^\circ]$. The natural frequency and damping ratio of each mode were computed as
\begin{equation}
    \omega_n = \sqrt{\mathrm{Re}(\lambda)^2 + \mathrm{Im}(\lambda)^2}, \qquad
    \zeta = -\mathrm{Re}(\lambda)/\omega_n.
\end{equation}

The model parameters were specified based on a representative swan goose configuration \cite{harvey2022birds}. The inertial and geometric properties were set as $m = 4.15\,\mathrm{kg}$, $I_{yy} = 0.081\,\mathrm{kg\cdot m^2}$, $AR_\mathrm{bird} = 7$, $S = 0.284\,\mathrm{m^2}$, and $\bar{c} = 0.163\,\mathrm{m}$. 
To account for tail contributions in the longitudinal dynamics, an equivalent horizontal stabilizing surface was included with projected area $S_h = 0.0180\,\mathrm{m^2}$ and moment arm between the gravity and the aerodynamic center of the tail $X_h = 0.261\,\mathrm{m}$.   
The same inertial parameters were used for both the avian-wing and airfoil-wing configurations to enable a consistent comparison. 

For each angle of attack, trim aerodynamic coefficients were obtained from wind tunnel measurements and used to determine the corresponding flight condition. Solutions yielding airspeeds outside a representative flight range ($10$–$30\,\mathrm{m/s}$) were excluded. Eigenvalues were then computed from the resulting linear system to assess stability.

\subsection*{Flow diagnostics}
\label{sub:method_flow}

\noindent
The flow analysis follows a consistent reduction framework to enable quantitative comparison across spanwise locations. 
Based on the 2D3C SPIV measurements, flow quantities are first evaluated in a chord-fixed coordinate system $(x_c,y_c,z)$, where $x_c$ is the chordwise coordinate, $y_c$ is the wall-normal coordinate, and $z$ is the spanwise coordinate (\autoref{fig:flow_analysis}E, \autoref{S_Fig:ExpSetup}). 
For each reported quantity, the local velocity fields are processed according to its specific definition, as described below, to obtain a chordwise-resolved quantity $\tilde{\phi}(x_c,z)$. 
These chordwise-resolved quantities are then reduced to representative section-level values by averaging over the valid chordwise measurement extent, thereby filtering local variability and extracting the dominant aerodynamic state of each section. 
Accordingly, the spanwise distribution is defined as
\begin{equation}
    \phi(z)=\frac{1}{L_c}\int_{L_c}\tilde{\phi}(x_c,z)\,\mathrm{d}x_c ,
    \label{eq:def_chordavg}
\end{equation}
where $L_c$ denotes the valid chordwise measurement extent. 
This reduction procedure is applied consistently to all reported quantities, including the edge-pressure coefficient ($C_{p,e}$), shape factor ($H$), peak turbulent kinetic energy ($k_{\mathrm{max}}$), wall-normal location of the TKE peak ($y_{\mathrm{peak}}$), Reynolds-stress anisotropy components ($b_{11}, b_{22}, b_{33}$), and dimensionless in-plane TKE production ($P_{\mathrm{plane}}^{*}$).

The boundary-layer edge velocity $u_e$ was identified from the chordwise velocity profiles in the chord-fixed coordinate system using a combined gradient and magnitude criterion \cite{schlichting2017boundary, vinuesa2016determining}. Specifically, the edge was defined as the first wall-normal location satisfying $\partial u_c/\partial y_c<0.02$ and $u_c>0.85\,u_{c,\max}$, where $u_c$ is the chordwise velocity component and $u_{c,\max}$ is the maximum chordwise velocity along the corresponding wall-normal profile.
The edge velocity is then used to estimate the edge-pressure coefficient via a Bernoulli approximation \cite{anderson2017fundamentals}, $\tilde{C_p} \approx 1 - (u_e/U_\infty)^2$. The boundary-layer edge location further defines the integration limit for evaluating the shape factor $\tilde{H}$ \cite{schlichting2017boundary}.

The turbulent kinetic energy is computed as $k=0.5(\overline{u'^2}+\overline{v'^2}+\overline{w'^2})$ \cite{pope2000turbulent}. At each chordwise position, the maximum value $\tilde{k}_{\mathrm{max}}$ and its wall-normal location $\tilde{y}_{\mathrm{peak}}$ are identified.

The Reynolds-stress anisotropy tensor is evaluated at $\tilde{y}_{\mathrm{peak}}$ as
\begin{equation}
    \tilde{b}_{ii}(x_c,y_{\mathrm{peak}},z)={\overline{u_i'^2}}/{2k_{\mathrm{max}}}-{1}/{3},
    \label{eq:bii}
\end{equation}
where $i=1,2,3$ correspond to the chordwise, wall-normal, and spanwise directions.

Mean-to-turbulence energy transfer is quantified using the in-plane resolved TKE production  \cite{pope2000turbulent}
\begin{equation}
    P_{\mathrm{plane}}(x_c,y_c,z)
    =
    -\sum_{i=1}^{3}\sum_{j=1}^{2}
    \overline{u_i' u_j'}
    \frac{\partial u_i}{\partial x_j}.
\end{equation}
Because only in-plane spatial gradients are available, $P_{\mathrm{plane}}$ represents the resolved contribution within the 2D3C measurement plane. $\tilde{P}_{\mathrm{plane}}^{*}(x_c,z)$ is then obtained from the wall-normal integral:
\begin{equation}
    \tilde{P}_{\mathrm{plane}}^{*}(x_c,z)
    =
    \frac{1}{u_\infty^3}
    \int_{0}^{\delta}
    P_{\mathrm{plane}}(x_c,y_c,z)\,\mathrm{d}y_c,
    \label{eq:P_plane}
\end{equation}
where $\delta$ is the boundary layer thickness.

\subsection*{Control cost analysis}
\label{subsec:method_control}

\noindent
To quantify how passive stability influences the need for active stabilization, a control-demand analysis was performed using a linear-quadratic regulator (LQR) framework \cite{kalman1960contributions, anderson2007optimal}. The longitudinal dynamics were augmented with a control input $\delta_e$ representing an equivalent elevator or tail deflection \cite{beard2012small},
\begin{equation}
    \dot{\mathbf{x}} = A\mathbf{x} + B\delta_e,
\end{equation}
with the optimal feedback law given by $\delta_e = -K\mathbf{x}$, obtained by minimizing a quadratic cost function.

For passively stable conditions, active stabilization was assumed unnecessary, and the control cost was set to zero. For unstable conditions, stochastic disturbances were introduced, and the required control effort was quantified from the closed-loop system response.

The primary metric reported in the main text is the average control-power coefficient $\overline{C}_{P,\mathrm{control}}$, which provides a non-dimensional measure of the control effort required to maintain stability. In addition, covariance-based quantities, including the state variance ($J_{\mathrm{state}}$), control input variance ($\sigma_{\delta_e}^2$), and control rate variance ($\sigma_{\dot{\delta}_e}^2$), were used to characterize the closed-loop response (\autoref{S_Fig:cost}).

\textbf{The resulting control-power coefficient should therefore be interpreted as a comparative proxy for stabilization demand, rather than a direct estimate of muscular or actuator energy expenditure.}
Full details of the control formulation, disturbance model, and covariance analysis are provided in the \autoref{supp:controlCostAnalysis}.

\newpage

\section*{RESOURCE AVAILABILITY}


\subsection*{Lead contact}

Requests for further information and resources should be directed to and will be fulfilled by the lead contact, Yang Xiang (\href{mailto:xiangyang@sjtu.edu.cn}{xiangyang@sjtu.edu.cn}).

\subsection*{Materials availability}

This study did not generate new materials.

\subsection*{Data and code availability}

\begin{itemize}
    \item All raw data reported in this paper will be shared by the \href{mailto:xiangyang@sjtu.edu.cn}{lead contact} upon request.
    \item All custom-made scripts and codes for analysis are available from the \href{mailto:xiangyang@sjtu.edu.cn}{lead contact} upon request.
    \item Any additional information required to reanalyze the data reported in this paper is available from the \href{mailto:xiangyang@sjtu.edu.cn}{lead contact} upon request.
\end{itemize}

\section*{ACKNOWLEDGMENTS}


The authors acknowledge Huibin Jiang, Zhuoqi Li, Wenchang Wang, and Yimin Wu for experimental support. This work was supported by the National Natural Science Foundation of China (Grant Nos. 12202273 and 91952302), the China Postdoctoral Science Foundation (Grant No. 2018M642007), and Shanghai Jiao Tong University’s ``Double First-Class'' Project.

\section*{AUTHOR CONTRIBUTIONS}

Conceptualization, L.C.; Methodology, L.C., S.Q., and Q.W.; Investigation, L.C.; Original Draft, L.C.; Review \& Editing, L.C., Q.W., J.H., Y.Y., and Y.X.; Funding Acquisition, Y.X., H.L., and S.Q.; Resources, Y.X. and S.Q.; Supervision, Y.X. and H.L.

\section*{DECLARATION OF INTERESTS}

The authors declare no competing interests.

\section*{DECLARATION OF GENERATIVE AI AND AI-ASSISTED TECHNOLOGIES}

During the preparation of this work, the authors used ChatGPT to improve language. After using this tool or service, the authors reviewed and edited the content as needed and take full responsibility for the content of the publication.

\section*{SUPPLEMENTAL INFORMATION INDEX}
\label{sec:supp}




\begin{description}
  \item Supplemental information of control cost analysis in a PDF.
  \item Figures S1-S5 in a PDF.
\end{description}

\bibliography{references}

\bigskip

\newpage

\newpage

\setcounter{section}{0}
\setcounter{figure}{0}
\setcounter{table}{0}
\setcounter{equation}{0}
\renewcommand{\thesection}{S\arabic{section}}
\renewcommand{\thefigure}{S\arabic{figure}}
\renewcommand{\thetable}{S\arabic{table}}
\renewcommand{\theequation}{S\arabic{equation}}

\section{Control cost analysis}
\label{supp:controlCostAnalysis}

\noindent
To quantify how passive stability alters the need for active stabilization, we constructed a comparative control-demand model based on a linear--quadratic regulator (LQR) \cite{kalman1960contributions, anderson2007optimal}. 
The purpose of this analysis is not to reproduce the full neuromuscular control strategy of a real bird or the complete actuator model of an aircraft, but to provide a consistent benchmark for comparing the active stabilization required by the avian-wing and airfoil-wing flyers under the same disturbance and control assumptions.

The uncontrolled longitudinal dynamics follow the linear state-space system introduced in the stability analysis,
\begin{equation}
    \dot{\mathbf{x}} = A\mathbf{x},
\end{equation}
with state vector
\[
\mathbf{x} = [\theta, u_\mathrm{f}, \alpha, q]^T ,
\]
where $\theta$ is the pitch angle, $u_\mathrm{f}$ is the forward-speed perturbation, $\alpha$ is the angle of attack, and $q=\dot{\theta}$ is the pitch rate. 
To represent active stabilization, a single control input $\delta_e$ was introduced, corresponding to an equivalent elevator or tail deflection \cite{beard2012small}. 
The controlled system is therefore written as
\begin{equation}
    \dot{\mathbf{x}} = A\mathbf{x} + B\delta_e .
\end{equation}
The control matrix was taken as
\begin{equation}
    B = [0,\,0,\,0,\,M_{\delta_e}]^T ,
\end{equation}
so that the control input acts through the pitching-moment equation. 
The dimensional control effectiveness is
\begin{equation}
    M_{\delta_e} = \frac{q_{\infty,0} S \bar{c}}{I_{yy}} C_{M_{\delta_e}},
    \label{eq:M_delta_e}
\end{equation}
where $q_{\infty,0}=\frac{1}{2}\rho u_{\infty,0}^2$ is the trim dynamic pressure, $u_{\infty,0}$ is the trim freestream velocity, $S$ is the reference wing area, $\bar{c}$ is the mean chord, and $I_{yy}$ is the pitch moment of inertia.

The pitching-moment control derivative was estimated using a conventional horizontal-tail volume approximation,
\begin{equation}
    C_{M_{\delta_e}} = -\eta_h a_h \tau_e V_h,
    \qquad
    V_h=\frac{S_h X_h}{S\bar{c}},
    \label{eq:tail_volume_control}
\end{equation}
where $V_h$ is the horizontal-tail volume coefficient, $S_h$ is the projected area of the equivalent horizontal stabilizing surface, and $X_h$ is the moment arm from the center of gravity to the aerodynamic center of the equivalent tail. 
Here $\eta_h$ denotes the tail dynamic-pressure efficiency, $a_h$ is the lift-curve slope of the equivalent horizontal tail, and $\tau_e$ is the elevator effectiveness factor. 
Because the present analysis is intended as a relative comparison rather than a detailed actuator-design study, the same equivalent tail geometry and control-effectiveness parameters were applied to both flyer configurations. 
In the calculations, $\eta_h$ and $\tau_e$ were set to unity, and the corresponding wing lift-curve slope was used for $a_h$. 
Thus, differences in the computed stabilization demand arise from the measured wing aerodynamics and the resulting flight dynamics. 
The representative values of $S_h$ and $X_h$ were specified from the swan-goose geometry used in the stability model \cite{harvey2022gull}.

The feedback law was obtained from a standard LQR formulation,
\begin{equation}
    \delta_e = -K\mathbf{x},
\end{equation}
where $K$ minimizes the quadratic cost
\begin{equation}
    J = \int_0^\infty \left( \mathbf{x}^T Q \mathbf{x} + \delta_e^T R \delta_e \right)\,dt .
\end{equation}
Unless otherwise stated, the weighting matrices were chosen as
\[
Q = I_{4\times4}, \qquad R = [1].
\]
This choice provides a consistent benchmark for cross-configuration comparison rather than a unique physically optimal tuning. 
Because the same state coordinates, weighting matrices, equivalent tail geometry, and control-effectiveness assumptions were used for both configurations, the resulting metrics should be interpreted as relative measures of stabilization demand.

For passively stable conditions, active stabilization was assumed unnecessary and the control cost was therefore set to zero. 
For unstable conditions, stochastic perturbations were introduced through the angle-of-attack state,
\[
G = [0,\,0,\,1,\,0]^T .
\]
The closed-loop system is
\begin{equation}
    \dot{\mathbf{x}} = A_{\mathrm{cl}}\mathbf{x} + G w,
    \qquad
    A_{\mathrm{cl}} = A-BK ,
\end{equation}
where $w$ is a zero-mean stochastic disturbance. 
The steady-state covariance matrix
\[
X = E[\mathbf{x}\mathbf{x}^T]
\]
was obtained from the continuous-time Lyapunov equation
\begin{equation}
    A_{\mathrm{cl}}X + X A_{\mathrm{cl}}^T + G Q_w G^T = 0,
    \label{eq:lyapunov_control}
\end{equation}
where $Q_w$ is the disturbance intensity. 
In the present comparison, the same disturbance intensity was used for all configurations; equivalently, $Q_w$ was set to unity after nondimensionalizing the disturbance amplitude. 
This formulation follows standard stochastic linear-control analysis \cite{anderson2007optimal}.

A control-power proxy was defined from the expected product of the control pitching moment and the pitch rate. 
The dimensional control moment is
\begin{equation}
    M_{\mathrm{control}}
    =
    q_{\infty,0} S \bar{c} C_{M_{\delta_e}}\,\delta_e .
\end{equation}
Because a stabilizing control moment may remove mechanical energy from the pitch motion depending on the sign convention, we used the magnitude of the mean control power as a non-negative comparative metric:
\begin{equation}
    P_{\mathrm{control}}
    =
    \left|E\!\left[M_{\mathrm{control}} q\right]\right|
    =
    \left|
    q_{\infty,0} S \bar{c} C_{M_{\delta_e}}
    E[\delta_e q]
    \right| .
\end{equation}
Using $\delta_e=-K\mathbf{x}$, this becomes
\begin{equation}
    P_{\mathrm{control}}
    =
    \left|
    q_{\infty,0} S \bar{c} C_{M_{\delta_e}}
    \left(-K\,E[\mathbf{x}q]\right)
    \right|,
    \label{eq:P_control}
\end{equation}
where $E[\mathbf{x}q]$ is the fourth column of the covariance matrix $X$. 
The corresponding nondimensional control-power coefficient is defined as
\begin{equation}
    \overline{C}_{P,\mathrm{control}}
    =
    \frac{P_{\mathrm{control}}}{q_{\infty,0} S u_{\infty,0}} .
    \label{eq:CP_control}
\end{equation}
This quantity is interpreted as a comparative indicator of active stabilization demand, rather than as a complete estimate of biological muscular expenditure or actuator energy consumption.

In addition to $\overline{C}_{P,\mathrm{control}}$, three covariance-based quantities were used to characterize the closed-loop response under stochastic forcing. 
The state variance
\begin{equation}
    J_{\mathrm{state}} = \mathrm{tr}(X)
\end{equation}
quantifies the overall fluctuation level of the longitudinal states. 
The control-input variance
\begin{equation}
    \sigma_{\delta_e}^2 = K X K^T
\end{equation}
measures the magnitude of the control action. 
To characterize actuation variability, we also estimated the control-rate variance. 
For this metric, the stochastic disturbance is interpreted as a finite-bandwidth perturbation with covariance $Q_w$, giving
\begin{equation}
    \dot{\delta}_e = -K\dot{\mathbf{x}}
    = -K(A_{\mathrm{cl}}\mathbf{x}+G w),
\end{equation}
and therefore
\begin{equation}
    \sigma_{\dot{\delta}_e}^2
    =
    K A_{\mathrm{cl}} X A_{\mathrm{cl}}^T K^T
    +
    K G Q_w G^T K^T .
    \label{eq:delta_dot_variance}
\end{equation}
This control-rate metric is used only as a complementary measure of actuation variability. 
Together, $\overline{C}_{P,\mathrm{control}}$, $J_{\mathrm{state}}$, $\sigma_{\delta_e}^2$, and $\sigma_{\dot{\delta}_e}^2$ provide relative measures of the stabilization burden across the avian-wing and airfoil-wing configurations.

\section{Supplementary figures}

\begin{figure}[htbp]
    \centering
    \includegraphics[width=0.7\textwidth]{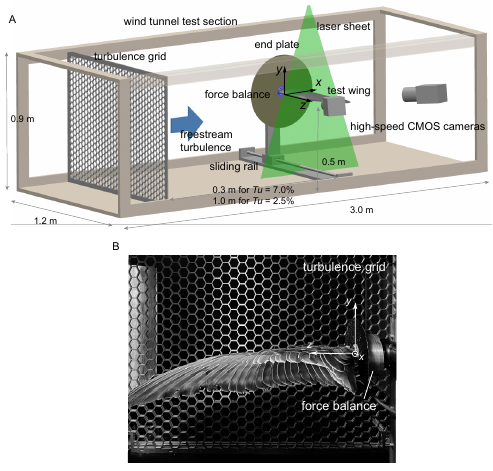}
    \caption{
    \textbf{Wind tunnel experimental setup.}
    (A) Schematic of the experimental configuration, including the turbulence generation system (grid), force measurement system (force balance, end plate, and sliding rail), stereoscopic PIV system (laser and cameras), and test specimen (wing model) \cite{qin2025flow}. 
    (B) Photograph of the experimental setup.
    }
    \label{S_Fig:ExpSetup}
\end{figure}

\begin{figure}[htbp]
    \centering
    \includegraphics[width=0.8\linewidth]{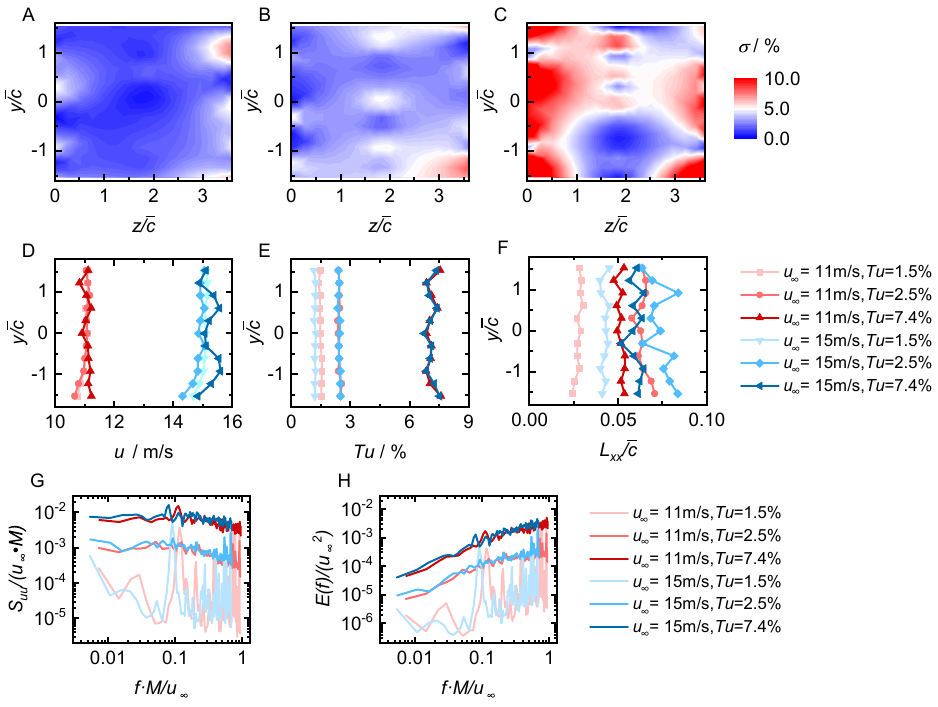}
    \caption{
    \textbf{Characterization of freestream turbulence conditions.}
    (A--C) Instantaneous velocity fluctuation fields in the $yz$-plane for $Tu = 1.5\%$, $2.5\%$, and $7.4\%$, respectively. 
    (D--F) Velocity profiles and turbulence characteristics measured at the spanwise mid-plane, including turbulence intensity $Tu$ and integral length scale $L_{xx}$. The integral length scale is included for reference and remains below $0.1\bar{c}$ \cite{thompson2023effects}.
    (G) Normalized power spectral density of velocity fluctuations.
    (H) Normalized turbulent kinetic energy distribution. 
    Three flow conditions were generated: $Tu = 1.5\%$ (no grid), $2.5\%$, and $7.4\%$, with the latter two obtained at streamwise positions $\Delta x = 1.0\,\mathrm{m}$ and $0.3\,\mathrm{m}$ from the grid, respectively, accounting for turbulence decay \cite{hearst2014decay,kitamura2014invariants}. The grid design follows Broatch \textit{et al.}'s work \cite{broatch2022automatized}.
    }
    \label{S_Fig:Tu}    
\end{figure}

\begin{figure}[htbp]
    \centering
    \includegraphics[width=0.8\linewidth]{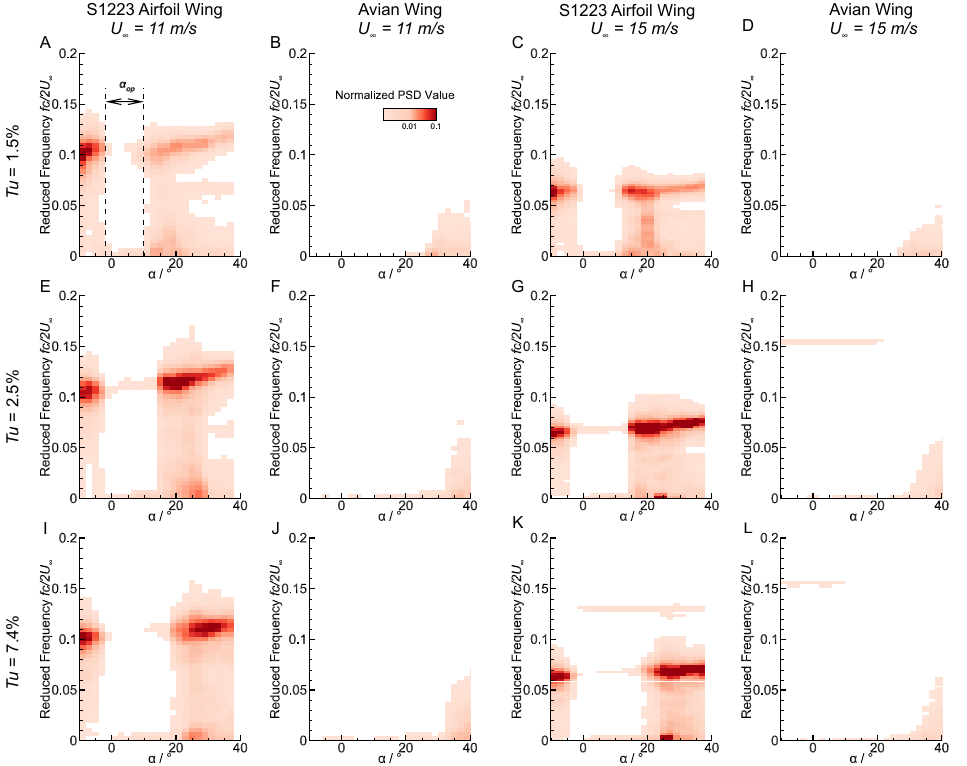}
    \caption{
    \textbf{Spectral characterization of lift fluctuations and operative angle-of-attack range.} Normalized power spectral density (PSD) of lift, $E_{C_L}u_\infty/\bar{c}$, is shown as a function of reduced frequency $f \bar{c}/2u_\infty$ for the avian wing and the S1223 airfoil wing at $u_\infty = 11$ and $15~\mathrm{m/s}$ over a range of angles of attack $\alpha$. Warmer colors indicate higher fluctuation intensity. Data with normalized PSD below $10^{-3}$ are omitted; the resulting blank regions indicate low-fluctuation conditions and are used to identify the operative angle-of-attack range $\alpha_{\mathrm{op,w}}$, complementing the definition based on lift-fluctuation magnitude in \autoref{fig:forceCoe_V11}D.
    }
    \label{S_Fig:liftForce_PSD}
\end{figure}

\begin{figure}[htbp]
    \centering
    \includegraphics{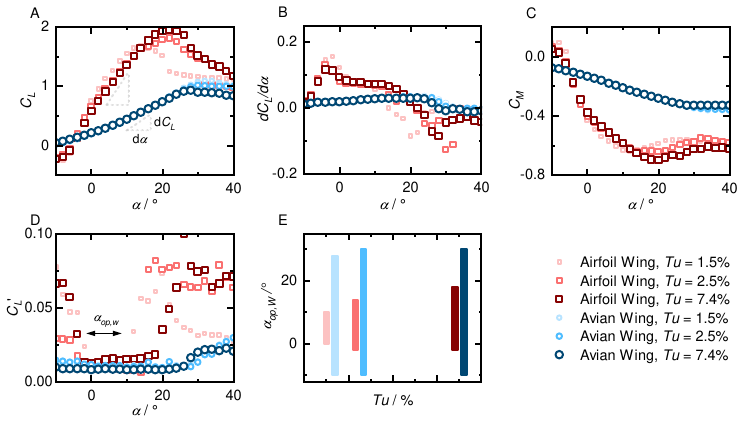}
    \caption{Aerodynamic characteristics of the swan-goose wing at $u_{\infty} = 15 \, \mathrm{m/s}$ ($Re = 1.65 \times 10^{5}$). The results demonstrate the avian adaptive aerodynamics, characterized by a broader operational angle of attack ($\alpha_{\mathrm{op,w}}$), reduced lift coefficients ($C_{L}$), shallower lift slopes ($\mathrm{d}C_{L}/\mathrm{d}\alpha$), and insensitivity to inflow turbulence. These performance trends are nearly identical to those observed at $u_{\infty} = 11 \, \mathrm{m/s}$ ($Re = 1.21 \times 10^{5}$) (\autoref{fig:forceCoe_V11}), underscoring the Reynolds-number-independence and consistency of the wing's passive adaptive mechanisms. }
    \label{S_Fig:force_swan_V15}
\end{figure}

\begin{figure}
    \centering
    \includegraphics{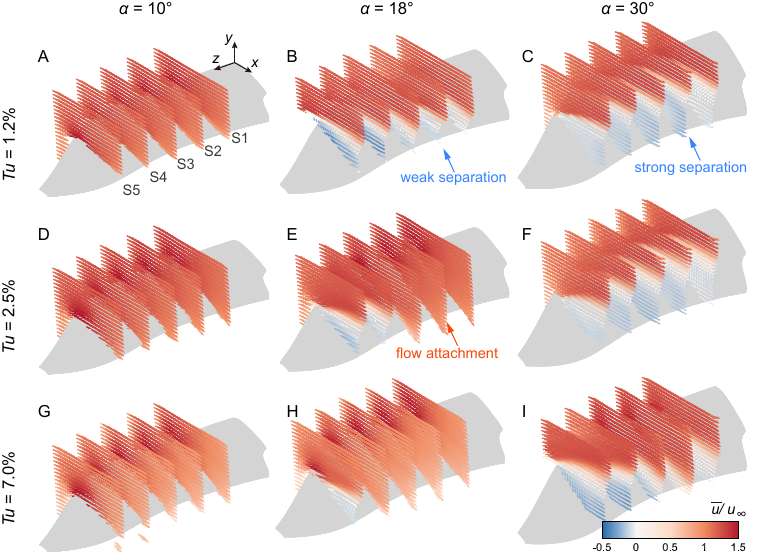}
    \caption{
    Flow separation over the suction surface of the airfoil wing. 
    The contours show the time-averaged streamwise velocity $\bar{u}$ normalized by the freestream velocity $u_\infty$, on representative chordwise sections. 
    Rows correspond to different freestream turbulence intensities, and columns correspond to angles of attack. 
    For cases with dense spanwise measurements (\autoref{fig:modelProblem}B), five representative sections are shown from the full set of fifteen measured sections to illustrate the spanwise evolution of the flow field. 
    Low-speed or reversed-flow regions, indicated by blue shades, provide a visual marker of boundary-layer separation. 
    }
    \label{S_Fig:flow_airfoil}
\end{figure}

\begin{figure}
    \centering
    \includegraphics{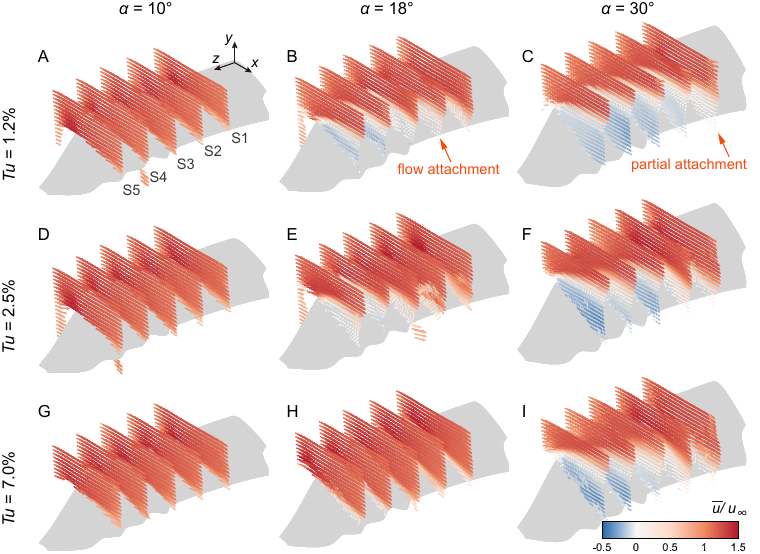}
    \caption{
    Flow separation over the suction surface of the avian wing. 
    Plotting conventions are the same as in \autoref{S_Fig:flow_airfoil}. 
    Compared with the airfoil wing, the avian wing retains more attached or partially attached flow across the low-turbulence conditions, indicating delayed and more localized separation.
    }
    \label{S_Fig:flow_avian}
\end{figure}

\begin{figure}
    \centering
    \includegraphics{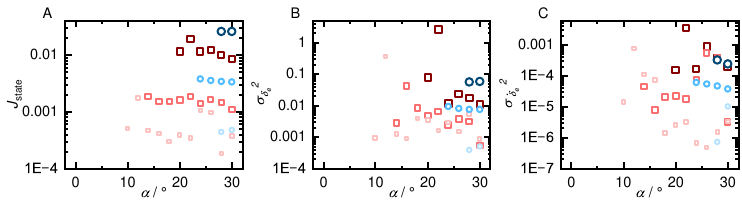}
    \caption{
    Covariance-based control metrics (derived in \autoref{supp:controlCostAnalysis}) under stochastic disturbances. 
    (A) State variance $J_{\mathrm{state}}$, 
    (B) Control-input variance $\sigma_{\delta_e}^2$, and 
    (C) Control-rate variance $\sigma_{\dot{\delta}_e}^2$ as functions of $\alpha$. 
    These quantities characterize the closed-loop response under stochastic forcing and provide complementary measures of control demand and actuation variability. The legend follows the same convention as in \autoref{fig:costAnalysis}.
    }
    \label{S_Fig:cost}
\end{figure}

\end{document}